\documentclass[12pt]{iopart}
\usepackage{iopams}
\expandafter\let\csname equation*\endcsname\relax
\expandafter\let\csname endequation*\endcsname\relax
\usepackage[fleqn]{amsmath}
\usepackage{amssymb}
\usepackage{amsbsy}
\usepackage{color}
\usepackage{graphicx}

\begin{document}

\title{A hybrid silicon-sapphire cryogenic Fabry-Perot cavity using hydroxide catalysis bonding}

\author{Yun-Long Sun, Yan-Xia Ye, Xiao-Hui Shi, Zhi-Yuan Wang, Chun-Jie Yan, Lei-Lei He, Ze-Huang Lu, and Jie Zhang}

\address{MOE Key Laboratory of Fundamental Physical Quantities Measurements, 
Hubei Key Laboratory of Gravitation and Quantum Physics, 
PGMF and School of Physics, Huazhong University of Science and Technology,
1037 Luoyu Road, Wuhan 430074, China}

\ead{yxye@hust.edu.cn, jie.zhang@mail.hust.edu.cn} \vspace{10pt}
\begin{indented}
\item[] December 13, 2018
\end{indented}

\begin{abstract}

The third-generation gravitational wave detectors are under development by operating the detector in cryogenic temperature to reduce the thermal noise. Silicon and sapphire are promising candidate materials for the test masses and suspension elements due to their remarkable mechanical and thermal properties at cryogenic temperature. Here we present the performances of the cryogenic thermal cycling and strength testing on hydroxide catalysis bonding between sapphire and silicon. Our results suggest that although these two materials have very different coefficients of thermal expansion, but if the flatness and the thermally grown $\mathrm{SiO_2}$ oxidation layer on the silicon surface are controlled well, the bonded samples can still survive thermal cycling from room temperature to 5.5 K. A breaking strength of 3.6$\pm 0.6$ MPa  is measured for the bonds between sapphire and silicon with a 190 nm silicon oxidation thickness after cooling cycle. We construct a hybrid sapphire-silicon Fabry-Perot cavity with the developing bonding technique in our lab. The measurement results reveal that the cavity can survive repeated thermal cycling while maintaining a good finesse.

\end{abstract}

\vspace{2pc} \noindent{\it Keywords}: hydroxide catalysis bonding, silicon, sapphire, cryogenic, ultra-stable cavity

\section{Introduction}\label{sec:intro}

Thermal noise is a major limiting factor both in gravitational wave interferometric detectors and ultra-stable lasers using  Fabry-Perot cavities \cite{cqg2000_Gonzalez, prl2004_Numata}. The state of the art strain detection limit for gravitational wave detectors is less than $\sim$$10^{-23}$, which can be improved by 1 to 2 orders with further reduction of temperature \cite{prl2016_Abbott, lrr2018_Abbott}. Therefore, future gravitational detectors such as the Einstein Telescope and the next generation advanced LIGO are planned to be operated at cryogenic temperatures to reduce thermal noise. At the same time, the Japanese gravitational wave detection project KAGRA has worked on the cryogenic temperature of about 20 K for several years and reach a strain less than $10^{-24}$ \cite{cqg2012_Somiya, prd2014_Hirose}. Monolithic suspension design for these new generation detectors has been required with hydroxide catalysis bonding (HCB) as the de facto standard. 

The HCB technique is first used in the Gravity Probe B telescope invented by Gwo \cite{spie1998_Gwo, patent2001_Gwo, patent2003_Gwo}. It has been used in Virgo, GEO 600 and LIGO projects to joint fused silica  \cite{cqg2010_Lorenzini, pla1998_Rowan, cqg2002_Willke, cqg2009_Smith, prt2018_Veggel}. Currently this bonding method is chosen to bond the sapphire parts in KAGRA for cryogenic temperature detection \cite{cqg2012_Somiya, prd2014_Hirose}. Silicon has also been proposed as low temperature test mass and suspension elements due to its excellent thermal and mechanical properties. There have been many reports of HCB of  silicon-silicon and sapphire-sapphire, but HCB of sapphire-silicon has received limited attention, especially the case under cryogenic conditions \cite{cqg2009_Veggel, cqg2011_Beveridge, phd2012_Beveridge, cqg2013_Beveridge, cqg2010_Dari, cqg2014_Douglas, aot2014_Veggel, cqg2015_Haughian, phd2015_Chen}.

In a related research field, the frequency stability of ultra-stable lasers is expected to reach $10^{-17}$ level when the reference Fabry-Perot cavities are operating at cryogenic temperature \cite{ prl2003_Muller, ol2014_Wiens, prl2017_Matei, prl2017_zhang}. To develop these cryogenic cavity lasers, highly reflective coating, high adaptability at low temperature, and well-assembled mirrors with lower mechanical loss are all significant factors to be considered in order to reach a smaller thermal noise limit \cite{prl2004_Numata, np2013_Cole}. All-sapphire cavity and all-silicon cavity have been reported before, but to our knowledge there has no report on hybrid sapphire-silicon cavity so far \cite{prl2003_Muller, ol2014_Wiens, prl2017_Matei, prl2017_zhang}.

Sapphire-silicon bonding presents a particular challenge due to their different coefficients of thermal expansion (CTE) from room temperature to cryogenic temperature \cite{jpc1983_Swenson, ht1983_White, ijt1997_White}. More importantly, silicon has a negative CTE from 124 K to 17 K compared with sapphire, which introduce a large stress during thermal cycling \cite{jpc1983_Swenson}. Here, we report an effective bonding between sapphire and silicon which can survive thermal cycling from room temperature to 5.5 K using an improved HCB technique. We manufacture several test pieces with different surface oxidize thickness silicon for bonding with sapphire, and test their properties through room temperature to cryogenic temperature cycling and bonding strength test, from which we obtain the most suitable technique for sapphire-silicon bonding. We believe this technique is important both for the development of gravitational wave detectors and cryogenic Fabry-Perot cavities.

With this bonding technique in hand, we build a 60-mm long hybrid cavity with sapphire mirror substrates and a silicon spacer. The cavity is able to survive several thermal cycling from room temperature to 5.5 K. The cavity finesse is measured to be larger than 5.2$\times10^5$ at 1070 nm after several cycling, proving the robustness of the designed and manufactured hybrid cavity at cryogenic temperature.

\section{Cavity design}\label{sec:CavityDesign}

We aim to design a cryogenic Fabry-Perot cavity which is transparent at 1070 nm. The fourth harmonic of this wavelength corresponds to the clock transition of the Al$^+$ ion optical clock that is under development in our group \cite{pra2017_Che, cpl2017_Zhang, ol2018_Zeng}. Sapphire is a suitable choice for mirror substrate due to its cryogenic properties and transparency at 1070 nm. However, silicon has a much smaller CTE at cryogenic temperature than that of sapphire. The CTE of silicon is expected to be zero at both 17 K and 124 K \cite{jpc1983_Swenson}. In addition, silicon has larger Young's modulus and smaller Poisson ratio, which makes it less sensitive to vibration. Therefore a hybrid sapphire-silicon cavity is supposed to be more stable than an all-sapphire cavity. Hence, we design a hybrid cavity with a 60 mm cuboid silicon spacer and two 25.4 mm diameter sapphire mirrors. 

The sapphire mirrors are coated with multiple $\mathrm{SiO_2/Ta_2O_5}$ layers, and the design working temperature is expected to be between 4$\sim$17 K. Because the CTEs of sapphire and silicon are quite different, the bond between silicon and sapphire is likely to break through the bonded surface when the cavity goes through thermal cycles. Before our work, there is no report demonstrating that the bond between silicon and sapphire can survive the thermal cycle from 300 K to low temperature. Therefore we prepare some samples of sapphire and silicon discs which are bonded with HCB method to investigate the true performance of the bonds. A number of crystalline silicon discs and sapphire discs of diameter 25.4 mm and thickness 6.35 mm are polished to nominal flatness of  $\lambda$/4. The normal orientations of polishing sides of sapphire and silicon are [1 1 1] and [0 0 0 1], respectively. The orientation is selected because the hybrid cavity is expected to have the best optical and mechanical properties through these directions \cite{prl2017_Matei}.

\section{Bonding and test of disc samples}\label{bonding}

\subsection{Oxidation and bonding}\label{subsec:oxidation}
\begin{figure}[!h]
\centering
\includegraphics[width=0.7\textwidth]{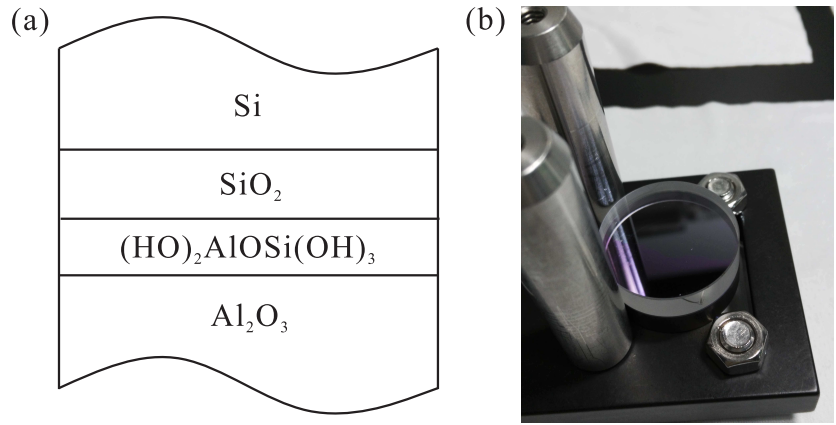}
\caption{(a) Layer structure of the bonding between silicon and sapphire. (b) Setup for the sample bonding. 
} \label{fig:layer}
\end{figure}

The layer structure of the HCB bonding between silicon and sapphire is shown in Fig. \ref{fig:layer} (a) \cite{cqg2009_Veggel, prd2016_Haughian}. The previous reported bonding technique between silicon and sapphire is quite robust at room temperature  \cite{cqg2010_Dari}. However, when we put two pairs of sapphire-silicon bonded samples with natural oxidation on silicon to cryogenic temperature, both of them break up during only one thermal cycle from room temperature down to 5.5 K, showing that the bonding strength with natural oxidation is not enough to overcome the internal stress during the thermal cycle. The HCB procedure commonly contains three steps: hydration and etching, polymerization, and dehydration. A layer of silicon oxide at the surface of silicon bulk is required for etching, and direct bonding with natural oxidation is demonstrated to be insufficient to survive thermal cycling. Thus, we choose dry thermal oxidation method with oxygen as reaction gas to make several test pieces with different oxidation thickness \cite{cqg2009_Veggel, cqg2013_Beveridge}. We then preform thermal cycling and strain test with the aim to find the most reliable method for cryogenic cavity bonding.

The silicon surface with natural oxidation is measured by an ellipsometer (Sentech Instruments, SENpro) and the measured oxidation thickness is about $4.0\pm 0.1$ nm. The oxidation thickness after dry thermal oxidation is measured with a stylus profiler (KLA-Tencor, D-120) by eliminating part of the oxide layer with hydrofluoric acid. To understand the effect of oxide thickness on the ultimate bonding performance, we prepare four sets of silicon samples (three pairs in one set) with 4 nm, 92 nm, 190 nm and 291 nm oxidation thickness for bonding with sapphire, respectively. 

Before bonding, we polish silicon discs and sapphire discs by cerium oxide powder and the discs are hydrophilic treated by sodium bicarbonate powder. We use deionized water for cleaning in between these steps, and wipe out the remaining water with methanol. Then the discs are placed in a home-built holder for bonding, as shown in Fig. \ref{fig:layer} (b). The bonding solution consists of sodium silicate solution (14\% NaOH and 27\% SiO$_2$) and deionized water with a volume ratio of 1:6, and it is filtered with 0.22 $\mu$m pore filters and deposited quickly on the center of the silicon surface with oxidation by a micro-pipettor. We choose the applied volume of bonding solution according to the bonding area, and it is approximately 0.4 $\mu$L/cm$^2$. Then sapphire disc is put on top of the silicon disc with a slight pressure. It is possible to align the discs in the beginning half minutes. Fully curing of the bonding at room temperature takes about four weeks.

\subsection{Thermal cycling}\label{ThermalCycling}

\begin{figure}[!h]
\centering
\includegraphics[width=0.7\textwidth]{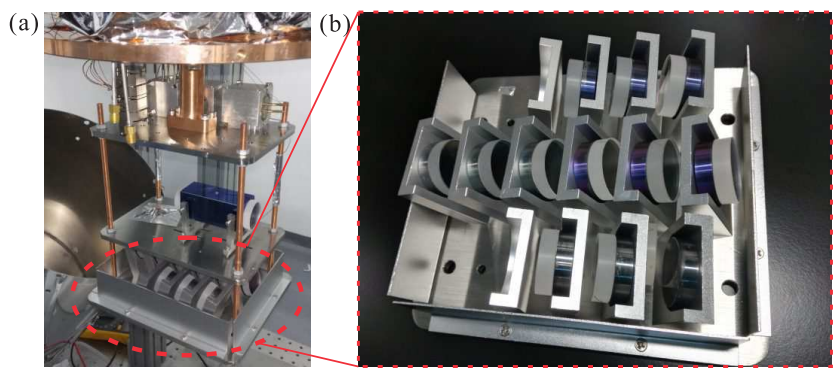}
\caption{(a) The experimental area for thermal cycle underneath a 4 K plate. (b) A top view of all samples.}
\label{fig:break}
\end{figure}

We then prepare different artificial oxidation layers on silicon and bond them with sapphire discs. We make twelve bonding samples with four different oxidation thicknesses, and three identical samples are prepared for each thickness. A home-built experimental area thermally contacted to a 4 K plate of a cryocooler (Cryomech, PT415) is used to provide the cryogenic environment, shown in Fig. \ref{fig:break} (a). Figure \ref{fig:break} (b) shows a top view of all samples. The U-shape sample holders are mounted to a plate with a 17$^\mathrm{o}$ tilt. The bonded samples settle themselves onto the holders due to the gravity. The lowest temperature of the holder plate during thermal cycle is measured to be 33 K due to the large thermal loads. This thermal cycling lasts 13 days with 10 days under 33 K. 

\begin{figure}[!h]
\centering
\includegraphics[width=0.7\textwidth]{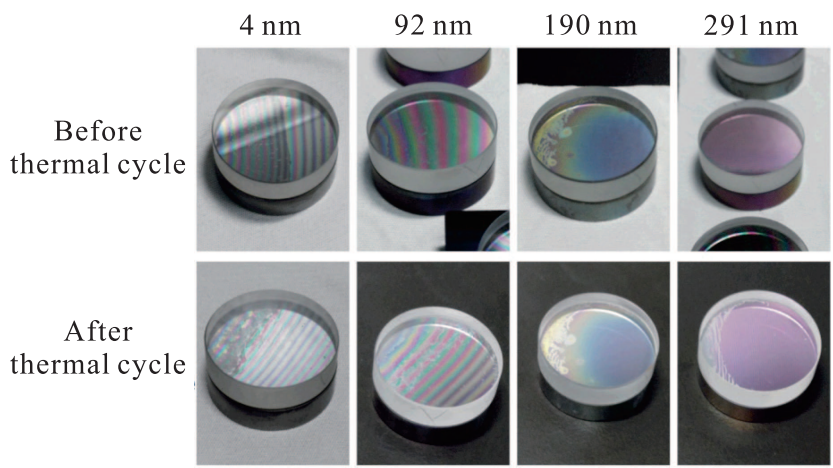}
\caption{Photos of the bonding samples before and after thermal cycle, from left to right, are 4 nm, 92 nm, 190 nm and 291 nm oxidation thickness, respectively.}\label{fig:samples}
\end{figure}

Figure \ref{fig:samples} shows the photos of the bonded samples before and after the thermal test. The different colors of the discs are due to different oxidation thicknesses, shown as gray, blue, and purple from thin to thick, respectively. Surface flatness of the discs and the bonding procedure affect the number and width of the interference fringes. After thermal cycle, one pair of bonding sample with a 4 nm oxidation thickness breaks apart while the other two pairs survive. All of the other nine pairs of oxide samples pass the thermal test. It shall be noted that the interference pattern of the survived samples change a little after thermal test, shown in Fig.~\ref{fig:samples}. 

In order to reach lower cryogenic temperature, we optimize the thermal contact of our experimental area, and decrease the temperature down to 11 K on the sample holder plate. We then test a second batch with four bonding sample pairs, and the oxidation thicknesses are 82 nm, 139 nm, 237 nm, 249 nm, respectively. All four pairs of bonded samples pass a thermal cycle for 22 days which contains 20 days at 11 K.  

We summarize the thermal cycling results of all bonded samples in Table \ref{thermalcycling}. All samples with artificial oxidation pass the thermal cycles. For samples without artificial oxidation, two samples split apart when cooled down to 5.5 K, and one of the three samples splits at the bonded surface during the cryogenic thermal cycle to 33 K. It is clear that the bonding samples without artificial oxidation is more likely to break during thermal cycles. However, samples with artificial oxidation layers (from 82 nm to 291 nm) demonstrate good thermal tolerance from 11 K to room temperature.

\begin{table}
\caption{\label{thermalcycling}\centering Summary of the thermal cycling results of all bonded samples.}
\footnotesize
\centering
\begin{tabular}{@{}cccc}
\br
Sample number&Oxidation thickness (nm)&Temperature (K)&Damage quantity\\
\mr
01, 02&3.9$\pm0.2$&5.5&2\\
03, 04, 05&3.9$\pm0.1$&33&1\\
06, 07, 08&92$\pm2$&33&0\\
09, 10, 11&190$\pm4$&33&0\\
12, 13, 14&291$\pm5$&33&0\\
15&82$\pm6$&11&0\\
16&139$\pm7$&11&0\\
17&237$\pm8$&11&0\\
18&249$\pm9$&11&0\\
\br
\end{tabular}\\
$^{*}$No. 01-05 are all natural oxidation samples.

\end{table}
\normalsize

\subsection{Breaking strength measurement}\label{subsec:breaking} 

Breaking strength is one of the most important parameters to characterize the bonding quality. A sample with large breaking strength means that it could work under larger stress and survive a longer time. There are mainly two measurements of breaking strength: tensile strength and shear strength. We choose to measure shear strength because of the horizontal hybrid Fabry-Perot cavity design. This design holds the mirrors on the orientation of shear strength. Therefore, we use the first test batch to carry out the shear strength test. 11 pairs of samples are left since one pair of natural oxidation sample splits during the thermal cycle. 

\begin{figure}[!h]
\centering
\includegraphics[width=0.8\textwidth]{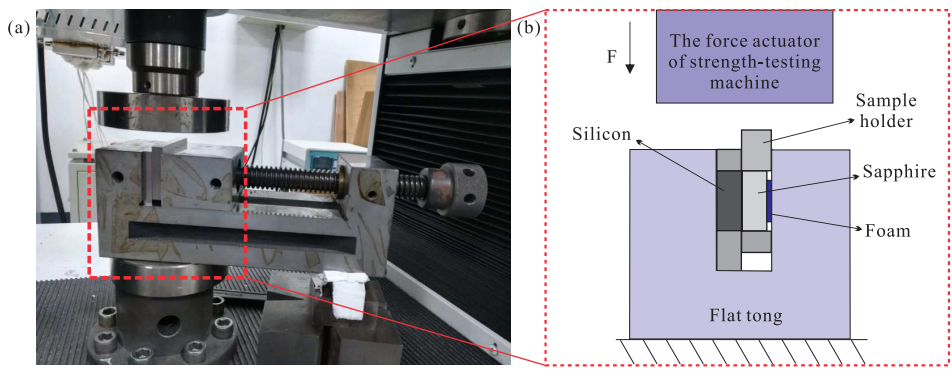}
\caption{ (a) Photograph of the strength test instruments. (b) Details of the sample holder and the test piece. }\label{fig:strengthtest}
\end{figure}

A photograph of the strength test is shown in Fig. \ref{fig:strengthtest} (a), and the details of the sample holder and the test piece are shown in Fig. \ref{fig:strengthtest} (b). To carry out the measurement, we use a precision universal tester (Shimadzu, AG-100kN), a flat tong, and two home-made stainless steel sample holders to hold the bonded sample. The universal tester applies a progressive force on the right sample holder, as shown in Fig. \ref{fig:strengthtest} (b). The flat tong is used to fix the left holder and make sure that the right holder can only move vertically when the progressive force is applied. All the interface surfaces of sample holders are polished to make sure fractionless tests. 

\begin{figure}[!h]
\centering
\includegraphics[width=0.7\textwidth]{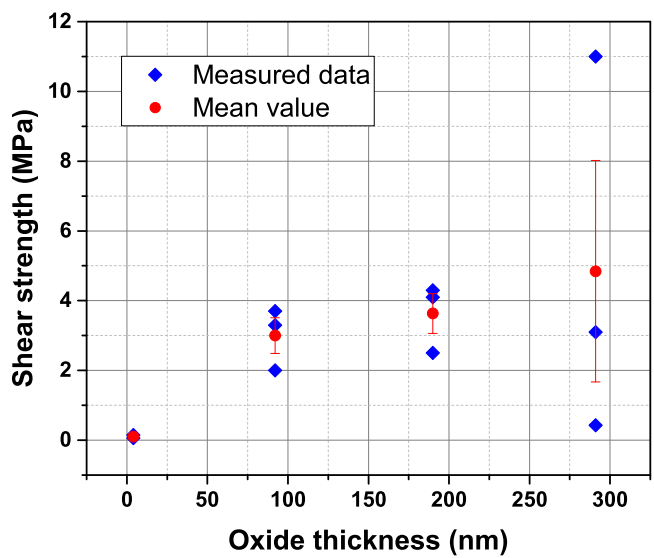}
\caption{ Measurement result of the breaking strength after thermal cycles for the four sets of bonds between silicon and sapphire.}\label{fig:shear}
\end{figure}

All the samples split through the bonded surface during the breaking strength measurement. The measured shear strength is shown in Fig. \ref{fig:shear}. The two surviving samples with the natural oxidation thickness have the lowest shear strength, which is $0.11 \pm 0.05$ Mpa. Meanwhile the other three sets of samples with artificial oxidation have higher shear strength, and the mean values are all higher than 3 MPa. The set of samples with 92 nm and 190 nm oxidation thicknesses have better shear strength performance and less deviation than that of the samples with 291 nm oxidation thickness. The large variation of the shear strengths of the three samples with 291 nm oxidation thickness might be induced by the changed surface flatness and randomly spread spikes \cite{cqg2009_Veggel}.

\begin{figure}[!h]
\centering
\includegraphics[width=0.7\textwidth]{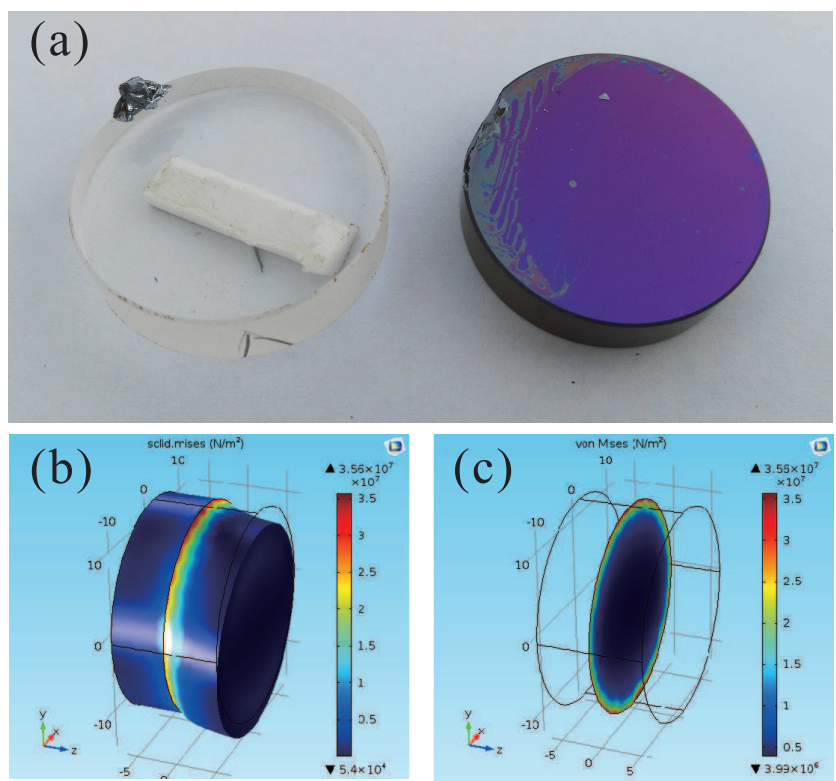}
\caption{ (a)  A typical photograph of the bonding sample after breaking strength test. (b) A simulated result shows the shear stress distribution using finite element analysis between silicon and sapphire cylinders. The left silicon cylinder is fixed, and a 2000 N force is applied on the right sapphire cylinder in the radial direction. (c) The stress distribution of the bonding surfaces.}\label{fig:shear1}
\end{figure}

It should be noted that most of the broken samples craze at the edges of the discs, as shown in Fig. \ref{fig:shear1} (a). Using finite element analysis, we analyze the stress distribution between a silicon cylinder and a sapphire cylinder. We build a geometric model consisting of the two silicon and sapphire cylinders, and fix the silicon cylinder. A 2000 N force is applied along the direction paralleled to the bond surface onto the sapphire cylinder. The simulation result is shown in Fig. \ref{fig:shear1} (b) and (c). It can be seen that the stress in the center area is almost the same and the stress on the edge is bigger than the stress in the center on the bond surface.

\section{Assembly and performance of the hybrid cavity}\label{sec:assembly} 

\subsection{Alignment and bonding of the cavity}\label{subsec:alignment} 

As introduced in Sec. \ref{sec:CavityDesign}, we choose a 60 mm long cuboid bulk silicon as the spacer. A pair of sapphire mirrors as the end mirrors is 25.4 mm in diameter and 6.3 mm thickness. The cross section of the spacer is a 25.4 mm$\times$25.4 mm square, the axial bore diameter is 8 mm, and the vent-hole diameter is 4 mm. The tolerance of the parallelity between the two end surfaces of the silicon spacer is about 3 millisecond. Due to the hardness of the silicon material, the flatness of two polished end surfaces is measured to be 0.40$\lambda$ and 0.43$\lambda$, respectively. According to discussion in Sec. \ref{bonding}, we choose a oxidation thickness about 100 nm for the silicon spacer, which is sufficient to pass the cryogenic temperature cycling and allows a reasonable breaking strength. The measured dry oxidation thickness for the two end surfaces is 123.7$\pm0.2$ nm and 123.2$\pm0.5$ nm, and the flatness is measured to be 0.52$\lambda$ and 0.56$\lambda$, respectively. The two commercial sapphire mirrors have a flatness of $\lambda$/10, and the mirrors are coated with a 5 mm diameter high-reflection coating on the central area. 

\begin{figure}[!h]
\centering
\includegraphics[width=0.7\textwidth]{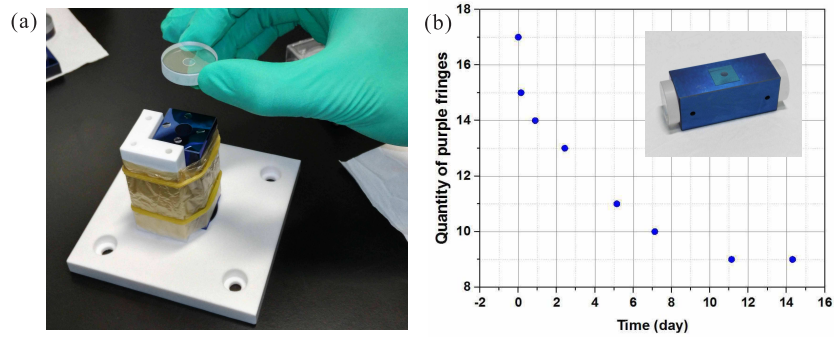}
\caption{(a) Align and bond of the mirrors of the hybrid cavity. (b) The number of the fringes changes with time. The inset is a photograph of the bonded hybrid cavity.}\label{fig:cavity}
\end{figure}

After surface processing, we use a home-made teflon adapter to align the sapphire mirror and the silicon spacer, shown in Fig. \ref{fig:cavity} (a). It helps the alignment once the mirror slides to touch the edge of the adapter. The yellow dust-free tape is used to avoid contamination of the spacer. One mirror is bonded to one end of the spacer by using the HCB method, and the other mirror has to be bonded by flipping the spacer four weeks later. During the curing period, we count the number of fringes on the bonded surface, as shown in Fig. \ref{fig:cavity}(b). The reduction of the fringe number is due to the continuous chemical reaction and the damping of the residual strain. The result indicates that the main visual chemical reaction takes about 12 days. 

\subsection{Finesse measurement}\label{subsec:Finesse} 
\begin{figure}[!h]
\centering
\includegraphics[width=0.7\textwidth]{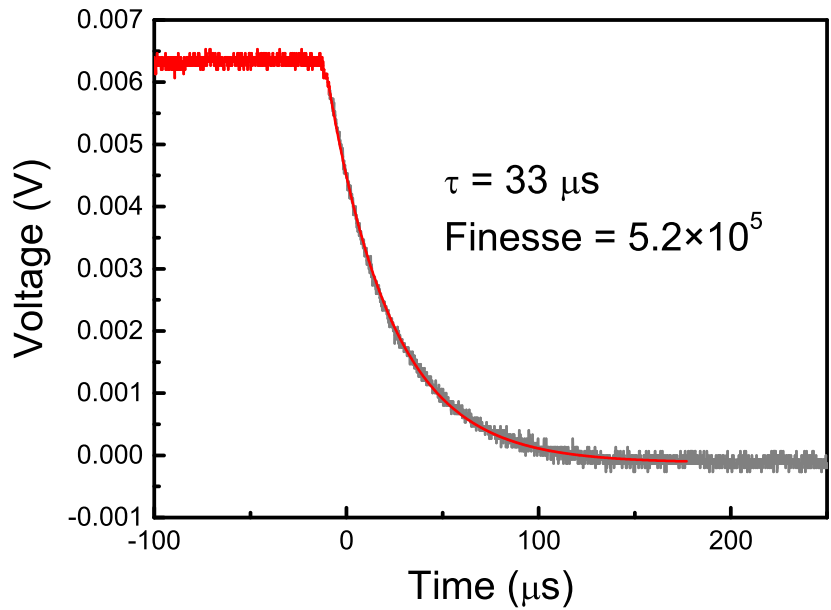}
\caption{Cavity finesse measurement at 5.5 K using the cavity ring down method.}\label{fig:finesse}
\end{figure}

In order to make sure that end mirrors are not contaminated during the HCB bonding procedure, we measure the cavity finesse using the cavity ring down method when the cavity is assembled. We house the bonded hybrid cavity into the cryocooler, and lock a diode laser onto its transmission peak by the Pound-Drever-Hall (PDH) technique \cite{apb1983_PDH}. The injected light is then chopped, and we monitor the decay of the cavity transmission signal to fit the ring down time and calculate the cavity finesse. One of the decay signals is shown in Fig. \ref{fig:finesse}. The cavity linewidth can be calculated to be $\Delta \nu = 1/2\pi\tau$. The finesse is $F={\Delta \nu_{FSR}}/{\Delta \nu}=5.2\times10^5$, where the free spectral range of the cavity is $\Delta \nu_{FSR}=c/2L$. We summarize all the measured cavity finesses under different conditions in Table 2. The first measurement is performed when the hybrid cavity is settled at room temperature, and the finesse of the cavity is 7.1$\times10^5$, which is a fairly good result in the state of art FP cavities. This demonstrates the feasibility of cavity construction using the HCB method, even though the complicated surface processing steps could easily pollute the area of the high reflection coatings. Then the cavity is cooled down to 20 K and 5.5 K in two different thermal cycles. The cavity passes the thermal cycles, and the finesse is 5.6$\times10^5$ and 5.2$\times10^5$, respectively. It takes longer than eight months between the first and the fifth measurement. During this time the cavity has experienced three thermal cycles and every thermal cycle is longer than one month. This implies that the cavity can not only tolerate repeated thermal cycling, but also maintain a good finesse.

\begin{table}
\centering
\caption{\label{finesse}\centering Measured finesses of the hybrid cavity at different temperatures.}
\footnotesize
\begin{tabular}{@{}lccc}
\br
Order&Temperature&Finesse/$\times10^5$&Linewidth/kHz\\
\mr
1st time&293 K&7.1&3.5\\
2nd time&20 K&5.6&4.4\\
3rd time&5.5 K&5.2&4.7\\
4th time&293 K&5.6&4.4\\
5th time&5.5 K&5.6&4.4\\
\br
\end{tabular}\\

\end{table}
\normalsize

\section{Conclusion}\label{conclusion}

In conclusion, we present an efficient method for assembling sapphire and silicon materials, which can survive cryogenic thermal cycling, and has a strong breaking strength.  We modify the traditional HCB method by adding an artificial oxidation layer to the silicon. Then we test the thermal stability of the bonded samples from room temperature to cryogenic temperature down to 5.5 K. We also perform the strength test indicating that the shear strength tolerance is larger than 3 MPa after thermal test using this bonding method. We further test the method by assembling a hybrid sapphire-silicon cavity, and obtain a finesse larger than  5.2$\times10^5$ after several cryogenic thermal cycles without contamination. The results of thermal cycling and strength testing in our experiments indicate that the modified HCB method meets the requirement of the third-generation gravitational wave detector in addition to the cryogenic ultra-stable laser systems.

\section{Acknowledgments} 
The authors wish to thank Liangcheng Tu, Pengshun Luo, Danhua Peng, and Jean-Michel Le Floch for the help of the oxidation of silicon and loan of the cryocooler. We also like to thank Jianhua Xiao of the State Key Laboratory of Material Molding and Mold for the measurement of the breaking strength of the samples. The project is partially supported by the National Key R\&D Program of China (Grant No. 2017YFA0304400), the National Natural Science Foundation of China (Grant Number 91536116, 91336213, 11774108, and 61875065).

\section*{References}

\end{document}